# Correlation of ground motion duration with its intensity metrics: A simulation based approach

Mojtaba Harati[1], Mohammadreza Mashayekhi[2]*, and Homayoon E. Estekanchi[2],

[1]Department of Civil Engineering, University of Science and Culture, Rasht, Iran
[2]Department of Civil Engineering, Sharif University of Technology, Tehran, Iran

**Abstract**

There are different kinds of intensity measures to characterize the main properties of the earthquake records. This paper proposes a simulation-based approach to compute correlation coefficients of motion duration and intensity measures of the earthquake ground motions. This method is used to investigate the influence of the ground motion data set selection in resulting duration-intensity correlation coefficients. The simulation procedure is used to tackle the problem of inadequate available ground motions with specific parameters. Correlation coefficients are investigated in three different cases. In case one, simulated ground motions differ in terms of earthquake source parameters, site characteristics, and site-to-source distances. In case two, ground motions are simulated in a specific site from probable earthquake events. In case 3, ground motions are simulated from a specific event in different sites. The first case doesn't show a significant correlation, while the second and the third case demonstrate significant positive and negative correlations, respectively.

## 1. Introduction

For seismic analysis of new or existing structures, seismic codes such as ASCE07 (2010) and rehabilitation provisions (e.g. ASCE/SEI 41-17 (2017) and FEMA-356 (2000)) typically recommend several frameworks, including Linear Static Procedure (LSP), Linear Dynamic Procedure (LDP), Nonlinear Static Procedure (NSP), and Nonlinear Dynamic Procedure (NDP). Each of this procedure has its own merit compared to the other one. For example, the LSP is so fast among the rest of the above-mentioned frameworks and can be readily used by a practitioner in engineering offices. However, it cannot include the nonlinear effects of the earthquake on the structural systems. On the other hand, the NDP is capable of considering nonlinearities which are raised both from materials and structural elements. While it is not as fast as linear frameworks and is a time-consuming process occasionally, the NDP incorporates the dynamic nature of the earthquakes and is thus taken to be the most reliable framework in the field of earthquake and structural engineering. Hence, NDP is used when structures with complex behavior are to be examined for an initial design or a procedure pertinent to structural retrofitting. These cases that need the NDP procedure may include base-isolated buildings or structures equipped with vibration control devices. For this kind of dynamic analysis, a set of ground motions should be taken and used as input load functions. In this case, a procedure called earthquake record selection is accomplished before any such an analysis. Earthquake records are collected according to the potential scenario earthquakes expected at the site. The scenario earthquakes are usually characterized by some significant parameters such as the magnitude of the possible earthquakes, source-to-site distance, local soil condition, duration of ground shakings, and some factors related to the amplitude-based intensity measures of the motions (Baker and Cornell 2006, Katsanos et al. 2010). In this regard, the peak ground acceleration (PGA), as well as spectral acceleration (SA) at the first structural period is considered as the amplitude-based intensity measures. These parameters are found to have good positive correlations with the structural damages quantified in the performance-based design framework or in the advanced dynamic analysis called IDA (Incremental Dynamic Analysis).

Record selection procedures are usually based on a combination of the above-mentioned parameters that define the characteristics associated with the expected scenario earthquakes. However, current seismic codes generally suggest a record selection process by which ground motions are chosen in such a way that their response spectrum is adequately compatible with a predefined target response spectrum (Katsanos et al. 2010). In this case, some rules are prescribed by the codes to confirm the aforementioned response spectrum compatibility. This compatibility is usually ensured in a way that the acceleration spectrum ordinates of the considered ground motion are adequately close to the values related to the target spectrum for a range of selected structural periods. It should be noted that the target spectra are either based on a design target spectrum or obtained from a probabilistic seismic hazard analysis.

Although the current design codes do not include the ground motion duration as the main selection criterion for the record selection procedure, a growing body of the research in this area shows that duration of the earthquake can have a significant impact on the structural responses. Several investigators have addressed the influence of motion duration on the structural responses. Their studies revealed that seismic responses of the structures under earthquake loadings with deteriorative behaviors, including RC frames (Belejo et al. 2017, Chandramohan et al. 2016, Han et al. 2017, Hancock and Bommer 2007, Raghunandan and Liel 2013), concrete dams (Sherong et al. 2013, Wang et al. 2015, Bin Xua et al. 2018, Wang et al. 2018) and masonry buildings (Bommer et al. 2004), are directly influenced by the duration of ground motions. It means that structures with deteriorating behaviors are much more susceptible to motion duration, so more structural and non-structural damages would be expected to happen at places whose constructions are exposed to long-duration ground shakings (Mashayekhi and Estekanchi (2012, 2013)). In this case, accumulated damage indices which are partially or completely based on the hysteretic cyclic energy of the earthquakes such as Pak-Ang damage index (Park and Ang 1985) are shown to have higher positive correlations with the motion durations. However, the extreme damage indices such as peak floor drifts or peak plastic rotations of the elements are demonstrated not to be well correlated to this parameter (Hancock and Bommer 2007, Sarieddine and Lin 2013, Mashayekhi et al. 2019). It is of the essence to note that the same results also apply for the steel (Bravo-Haro and Elghazouli 2018, Chandramohan et al. 2016, Kiani et al. 2018) and wood frame (Pan et al. 2018) structures.

There are more than 30 definitions for motion duration in the literature, but some of them are more commonly accepted and used by the earthquake engineering community. Among the defined available definitions in the literature, bracketed duration, uniform duration as well as the significant duration are more repeatedly used in the field of earthquake engineering. The bracketed duration of motion delivers the total time left between the first and last acceleration excursions which are greater than a specific predefined threshold. The definition pertinent to the uniform duration is all related to the sum of the elapsed time intervals considering the same aforementioned threshold level set on the acceleration (Bommer and Marytínezpereira 1999). But the definition related to significant duration is somehow different from the bracketed and uniform duration. This definition of the motion duration takes use of a well-known integration-based accumulative intensity measure, the so-called Arias Intensity (AI). Significant duration is denoted by $D_{x-y}$ hereafter, which is defined as the time interval during which the normalized AI moves from a minimum (x%) to a maximum (y%) threshold. And so, the $D_{5-75}$ means the time interval as buildup accumulation energy of the earthquake goes up from 5 to 75 percent. It is worthwhile mentioning that other thresholds for different applications have been also selected for the definitions of significant duration in the literature so far, which are denoted as $D_{5-95}$, $D_{20-80}$ and $D_{15-85}$ as well. It should be noted that the AI of a ground motion may get altered when the acceleration time function is changed or scaled, but its related significant duration remains unchanged altogether. Therefore, contrary to the definition of AI that depends on both motion duration and amplitude-based intensity measure, the significant duration is completely dependent on the duration of motions and treated as a duration-related intensity measure. It is of the essence to add that some studies show that the Cumulative Absolute Velocity (CAV), which is defined in the next section, can be also considered as an alternative for the AI to assess the effect of the motion duration on structural responses (e.g., EPRI 1988, Cabañas et al. 1997). This is due to the fact that both of these intensity measures, the CAV and AI, are capable of capturing and showing the cumulative energy of the ground motions.

Few researchers in the past focused on exploring the correlations of duration-related intensity measure with the amplitude-based intensity metrics—for instance, the SA, PGA or PGV—and the CAV as a cumulative intensity parameter. Bradley (2015) used a combination of ground motion prediction equations (GMPEs) and the bootstrap sampling method to find the involved correlation coefficients for different intensity measures. He points out that a high positive correlation exists between AI and CAV, but he fails to address such a correlation between CAV and significant duration. Moreover, it is demonstrated that a good positive correlation is found between SA and AI, not with significant duration, over a range of short structural periods while these two considered parameters are not well correlated in periods elsewhere. Bradley (2012) and Baker and Bradley (2017) computed linear correlation coefficients for examining such correlations from the observed data of NGA projects. They found that significant duration is negatively correlated with SA in points located in the range of short to medium structural periods although a low positive correlation can be observed in the zone related to long periods of vibration. Bradley (2012) also reported a low correlation between the CAV and significant duration. In this case, he took the $D_{5-75}$ parameter as the metric for the significant duration and concluded that they, the CAV and the $D_{5-75}$ parameter, are poorly correlated to each other.

He also found a strong correlation between the $D_{5-75}$ and $D_{5-95}$ duration-related parameters, so it can be deduced and generalized from this research that the CAV and significant duration are not well correlated overall. Mashayekhi et al. (2019) studied the governed relationship of duration-related parameters, the CAV and significant duration, and the ones related to the amplitude-based intensity measures—the PGA and SA. They found that motion duration and the amplitude-based intensity measures are correlated and convoluted to each other through an exponential function. They then applied their findings into nonlinear structural assessment and found that considering motion duration can have a strong impact on structural performance. The correlation of SA and the effective number of nonlinear cycles as a duration-related intensity parameter has also been explored by yet another research (Du and Wang 2017). It was reported that these two earthquake intensity metrics have a moderately good correlation in short structural periods while they experience a descending trend for the correlation coefficient values computed for the medium to long periods of vibration.

In this paper, a new method for computing correlations between duration-related parameter—or significant duration of ground motion—and the ones pertinent to the cumulative and amplitude-based intensity measures is proposed. In this method, a simulation is conducted using Monte Carlo data sampling on the data provided by the selected attenuation relationships, which makes possible the calculation of correlation coefficients between duration and the selected intensity measures. Simulated events provide continuous data by which statistical analysis can be more accurately accomplished. More, all probable values for the involved variables are covered in a most effective way. With the aid of the simulation procedure, different conditions for the scenario earthquakes can be also modeled for the calculation of correlation coefficients. In this case, selective conditions for the scenario earthquakes can be produced through the proposed simulation method, whereas the results found by the real collected motions are restricted to the condition existed in the selected database. For the rest of the paper, the hired intensity measures are introduced first, and then the simulation and the methodology used in this study for figuring out the correlation coefficients of duration and other intensity measures are discussed thoroughly. Next, the results associated with the presented numerical examples of three different conditions are examined in detail. Finally, a discussion section would be provided in order to elaborate on the matters related to the computed correlations in different simulation conditions, their implications with the record selection procedure.

## 2. Considered intensity measures

In this investigation, the PGA, PGV, spectral acceleration (SA) or the pseudo-spectral acceleration (PSA) at different structural periods have been taken as the amplitude-based intensity measures. The PGA and PGV are the peak values of the earthquake time series, which are related to the acceleration and velocity profile of the motion respectively. The amplitude-based intensity measures are commonly accepted and used as the major record selection criteria in the PBD framework. They are also regarded as the main intensity measures required in the IDA and nonlinear time history response analysis. On the other hand, the significant duration is taken as the duration-related intensity measure. While there are many definitions for the motion duration in the literature, the definition for the significant duration is selected as a duration-related parameter because reliable attenuation relationships, by which simulation procedure can be readily carried out, has been developed for this duration definition. The procedure pertinent to the calculation of a form of significant duration, the $D_{5-75}$ parameter, for the Loma-Prieta earthquake of 1989 is depicted in Figure 1. According to the figure, the significant duration is the time interval during which the buildup energy of the normalized AI moves from a minimum (5%) to a maximum (75%) threshold. The times associated with the mentioned minimum and maximum thresholds are defined by $t_x$ (here 8sec) and $t_y$ (here 12.8sec), respectively. To evaluate the strong motion duration, it is reported that the CAV can be used interchangeably with the AI in the definition of significant duration (e.g., EPRI 1988, Cabañas et al. 1997).

Both of the CAV and AI are defined as the time integral of a form of acceleration function profile as can be seen in Equation (1) and (2), where the $|a(t)|$ is the absolute value of the acceleration function of the ground motion at time t, [a(t)]. Also, $t_{max}$ and AI is the total duration of ground motion and the total AI calculated for the entire duration of the ground shakings. It can be readily understood from the given form of the above-mentioned equations that both of these intensity measures increases with time and have the capacity to capture the accumulative characteristics of the earthquakes. This ability is in marked contrast to what can be grasped by the amplitude-based intensity measures, such as the PGA, PGV and SA.

$$AI = \frac{\pi}{2g} \int_0^{t_{max}} [a(t)]^2 \, dt \qquad (1)$$

$$CAV = \int_0^{t_{max}} |a(t)| \, dt \qquad (2)$$

It should be noted that the AI of a ground motion may get altered when the acceleration time function is changed or scaled, but its related significant duration remains unchanged altogether. Therefore, contrary to the definition of AI, the significant duration is completely treated as a duration-related intensity measure. This is due to the fact that the significant duration is mainly a function of motion duration, and it is not changed when the amplitudes of a ground motion record or its related response spectrum are altered. The CAV is also similar to the AI in this case because its definition is dependent on both motion duration and amplitude-based intensity measure. Consequently, AI and CAV can be actually considered as the cumulative intensity measure of the earthquakes. In this study, the CAV is just hired as the cumulative intensity measure, not as a duration-related metric.

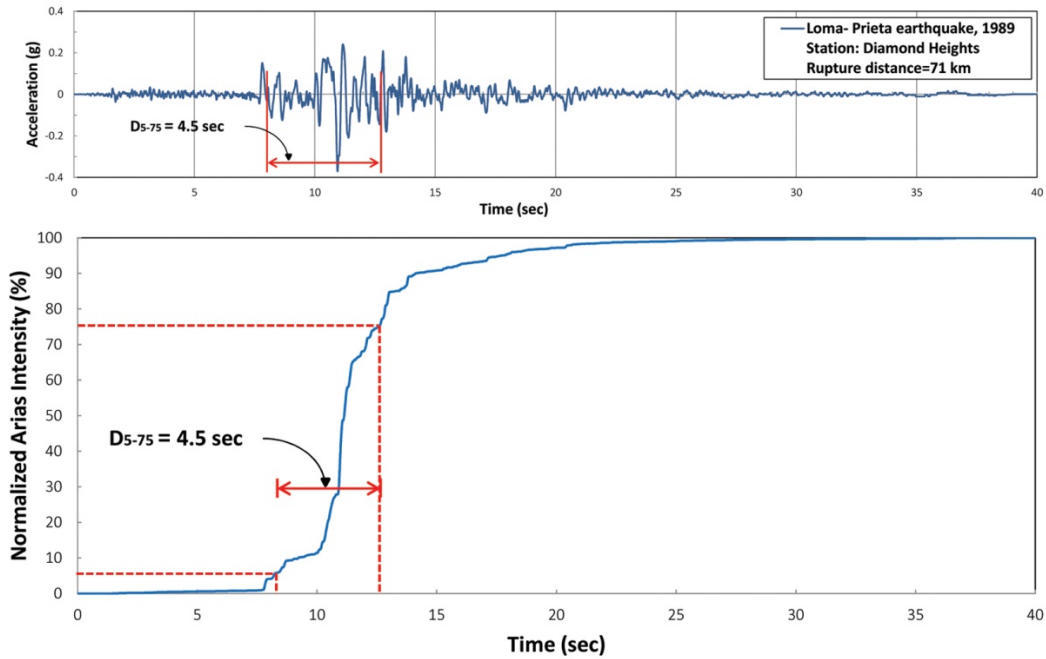

Figure 1. The procedure required to compute the $D_{5-75}$ parameter of a recorded ground motion

### 3. Monte Carlo Simulation

Monte Carlo Simulation (MCS) is a powerful statistical method which has been applied to many engineering problems across different disciplines. Researchers and engineers use this method to conduct virtual experiments on the computer. MCS is employed in complex and highly nonlinear engineering models, for it can deal with a lot of random variables that have different distribution types. For each experiment, a set of input random variables $I = (I_1, I_2, \ldots, I_n)$ is sampled or generated, which are based on their related distribution functions. Then the output variable $Y$ is computed using a performance function $S = g(I)$ with the input values randomly generated for each round of the experiment. While a lot of experiments like this should be conducted in the MCS procedure, a set of samples for output variable $Y$ is produced and then become available for statistical analysis and interpretation.

In this paper, the MCS technique is taken for carrying out a computer simulation, allowing a new scheme through which the calculation of correlation coefficients between two intensity measures would be possible. One of these intensity measures would be always the duration of motion, the significant duration parameter. On the other hand, the cumulative or amplitude-based intensity measures are adopted as the opposite variable of motion duration. In this case, several desired conditions for the scenario earthquakes can be modeled through the experiments done with the

MCS process, which let us examine the correlation of motion duration against the cumulative and amplitude-based intensity measures. While the inadequate amount of recorded data for large earthquake events makes computational problem in statistical analysis, especially for large intensity measures, MCS procedure is performed to such an extent that the number of events at the different level of intensity measures is nearly the same. This is because the confidence interval length has an inverse relation with the number of samples, so providing an equal number of events for all possible intensity measures is the main advantage and suitability of this method.

## 4. Methodology

### 4.1. Characteristics of the simulation model

Using the MCS procedure, the proposed method can offer a framework to compute correlation coefficients between duration-related parameter versus the cumulative and amplitude-based intensity measures. In this regard, thousands of possible earthquake scenarios are simulated by data sampling, which is totally in contrast to the use of real ground motions that are limited to the finite number of previously recorded motions. The main advantage of using simulated data is that it is possible to seek the correlation of a duration-related parameter and the other intensity measures with sufficient amounts of data, especially for the higher levels of intensity measures.

For the simulation procedure, all possible values of the duration-related parameter, amplitude-based as well as cumulative intensity measures are sampled and determined. While there are not an ample number of recorded ground motions with the same specific condition (e.g. with the same source-to-site distance or exactly with same M, moment magnitude), for each scenario, intensity measures and duration-related parameter are computed by existing GMPEs. As mentioned in the preceding sections, in this study, PGA, PGV and 5% damped acceleration spectrum are nominated to be simulated as the amplitude-based intensity parameters. For these intensity measures, the equations developed by Campbell and Bozorgnia (2014) are employed hereafter because we assume that it is more relevant to the selected sites. The significant duration ($D_{5-75}$) is also chosen to be at work as the duration-related parameter. In this case, the attenuation equation developed by Afshari and Stewart (2016) is taken to be employed for the calculation of the $D_{5-75}$ parameter. These equations need a number of parameters as their input variables coming in the following paragraphs.

Nearly all attenuation relationships need a parameter related to the soil condition, the so-called VS30. This is the time-averaged shear wave velocity over a sub-surface depth 30 meters. In the simulation process, the Z1.0 and Z2.5 are the depth parameters and are defined as the depth level at which shear wave velocity reach 1000 m/s and 2500 m/s, respectively. The Z1.0 depends on the VS30 and is calculated according to a relationship developed by Abrahamson and Silva (2008) as expressed by Equation (3). The Z2.5 is then computed by an extrapolation procedure based on Z1.0 parameter as recommended by Campbell and Bozorgnia (2006). One of the other parameters that should be defined for the simulation model is the source-to-site distance parameter, the rupture distance ($R_{rup}$). This parameter is defined as the slant distance to the closest point on the rupture plane.

$$Z_{1.0} = \begin{cases} \exp(6.745) & V_{S30} < 180 \, m/s \\ \exp\left[6.745 - 1.35\ln\left(\frac{V_{S30}}{180}\right)\right] & 180 \leq V_{S30} \leq 500 \, m/s \\ \exp\left[5.394 - 4.48\ln\left(\frac{V_{S30}}{500}\right)\right] & V_{S30} > 500 \, m/s \end{cases} \quad (3)$$

In addition to rupture distance, existing attenuation relationships may also need the Joyner-Boore distance ($R_{JB}$) which is defined as horizontal distance to the surface projection of the rupture. This distance is independent of the rupture distance in general. Using a vertical cross-section through a fault rupture, a plane schematic illustration of earthquake source and distance measures is shown in Figure 2.

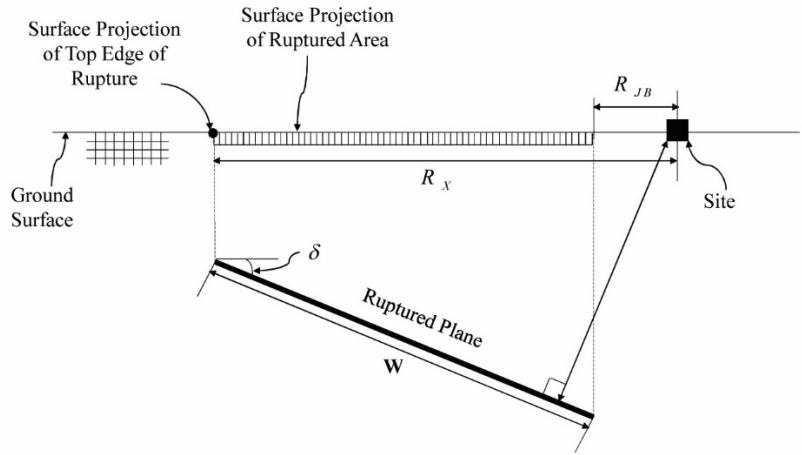

**Figure 2. Schematic illustration of earthquake source and distance measures using a vertical cross-section through fault rupture plane (Kaklamanos et al. 2011)**

It is worthy to add that three fault types are considered, namely the normal, reverse and strike-slip. In this figure, $\delta$ is fault dip, W is down-dip rupture width, and $Z_{TOR}$ is depth-to-top of the rupture. The dip is the angle that a planar geologic surface is inclined from the horizontal one, where it is assumed to be vertical in the strike-slip faults ($\delta$=90). Moreover, the average values of dip angle equal to 50 and 40 are recommended for normal and reverse faulting events, respectively (Kaklamanos et al. 2011). $R_X$ is the horizontal distance to the surface projection of the top edge of the rupture, which is measured perpendicular to the fault strike and is computed by Equation (4). In this equation, $\alpha$ is the source to site azimuth that for a given site is the angle between the positive fault strike direction and the line connecting the site to the closest point on the surface projection of the top edge of the rupture (Chiou 2005). This angle is assumed positive when it is measured clockwise as shown in Figure 3.

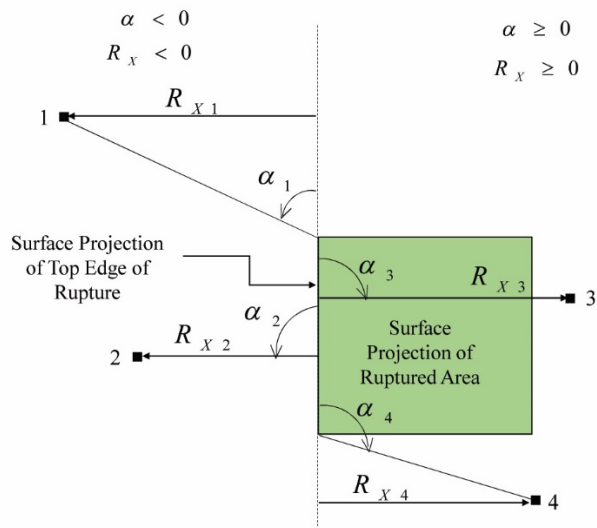

**Figure 3. Plan view of a fault rupture (Kaklamanos et al. 2011)**

$$R_X = \begin{cases} R_{JB} |\tan \alpha| & 0 \leq \alpha < 90 \text{ and } 90 < \alpha \leq 180, \ R_{JB} |\tan \alpha| \leq W \cos \delta \\ R_{JB} \tan \alpha \cos\left[\alpha - \sin^{-1}\left(\dfrac{W \cos \delta \cos \alpha}{R_{JB}}\right)\right] & 0 \leq \alpha < 90 \text{ and } 90 < \alpha \leq 180, \ R_{JB} |\tan \alpha| > W \cos \delta \\ R_{JB} + W \cos \delta & \alpha = 90, R_{JB} > 0 \\ \sqrt{R_{RUP}^2 - Z_{TOR}^2} & \alpha = 90, R_{JB} = 0, R_{RUP} < Z_{TOR} \sec \delta \\ R_{RUP} \csc \delta - Z_{TOR} \cot \delta & \alpha = 90, R_{JB} = 0, R_{RUP} \geq Z_{TOR} \sec \delta \\ R_{JB} \sin \alpha & -180 \leq \alpha < 0 \end{cases} \quad (4)$$

The relationship developed by Wells and Coppersmith (1994) is used to estimate the down-dip rupture width (W) from moment magnitude and the style of faulting as brought in Equation (5).

$$W = \begin{cases} 10^{-0.76+0.27M} & \text{for strike-slip events} \\ 10^{-1.61+0.41M} & \text{for reverse events} \\ 10^{-1.14+0.35M} & \text{for normal events} \end{cases} \quad (5)$$

The method employed by Kaklamanos et al. (1994) is used to estimate the depth-to-top-of-rupture ($Z_{TOR}$) from hypocentral depth ($Z_{HYP}$), down-dip rupture width (W), and dip angle ($\delta$) as expressed in Equation (6). The Hypocentral depth is also computed according to Equation (7).

$$Z_{TOR} = \max\left[(Z_{HYP} - 0.6W \sin \delta), 0\right] \quad (6)$$

$$Z_{HYP} = \begin{cases} 5.63 + 0.68M & \text{for strike-slip faulting} \\ 11.24 - 0.2M & \text{for non-strike slip faulting} \\ 7.08 + 0.61M & \text{for general (unspecified) faulting} \end{cases} \quad (7)$$

### 4.2. Variable selection for the considered simulation cases

In this study, three types of ground motion data set are considered to be the subjects of simulation. In order to generate each type of ground motion data set, a distinct simulation procedure is defined. Simulation procedures are designed so that generated motions have the desired characteristics. Based on the types of sampling, the involved variables (which are explained in the previous section) are determined for each simulation scenario. Each simulation is modeled and accomplished according to five independent variables, where according to what case they belong to, they can be either a constant quantity or a random variable. Then these variables are sampled and generated to build the characteristics of the scenario earthquakes. They include parameters related to the source points (Fault type and the M), the property or soil condition of the site to where earthquakes are received (VS30) and variables pertinent to the path through which earthquakes travel to reach the site of interest ($R_{rup}$ and $R_{JB}$). Three different cases, representing different ground motion (GM) data sets, by which modeling of the simulations are conducted are as follow:

- **Case 1-Random GMs:** for this simulation case, events are produced in a way similar to the situation where they belong to a large database composed of collected real data. As expected, earthquakes in such a database have different and miscellaneous characteristics, including different soil conditions, fault types and source-to-site distances. Therefore, in this simulation case, all involved variables are modeled using random variables with uniform distributions.

- **Case 2-Same site GMs:** in this simulation example, earthquakes are generated in such a way that is similar to assuming the situation where a specific site with a fixed seismometer in a location is considered. In this case, many earthquakes are recorded by the seismometer during its lifespan. As a result, the moment magnitude, M, is modeled with a random variable of a uniform distribution. And the rest of the parameters are considered as constant terms. As can be readily concluded, this simulation procedure may seem to be in contrast to the nature of the earthquake phenomenon in which larger events are rare to happen, which is also justified by the Gutenberg–Richter law. However, a large number of simulations in high seismic levels can enhance the reliability of our statistical analysis. Since the location at which ground motions are recorded has a specific soil condition, the time-averaged shear wave velocity over a sub-surface depth 30 meters (VS30) is assumed to be constant. The source-to-site and the Joyner-Boore distances ($R_{rup}$ and $R_{JB}$) are employed to be as a constant variable for this case because there is regularly a constant distance between the faulting point and where ground motions are recorded. The Joyner-Boore distance is independent of the rupture distance in general, but in this simulation scenario, the $R_{JB}$ is assumed to be equal to 1/3 of the rupture distance.

- **Case 3-Same source GMs:** for this simulation category, earthquakes are first generated at the fault location and then propagated through the layers of the earth. This represents a situation in which an earthquake, which is happened and traveling from its source, is recorded by many seismometers located at different distances from the hypocenter. Thus, the source-to-site and the Joyner-Boore distances ($R_{rup}$ and $R_{JB}$), as well as the VS30 parameter, are modeled using random variables with uniform distributions for this simulation case of interest. It is worth to add that the considered range of $R_{JB}$ variable is selected in a way that its generated values would be equal to 1/3 of those related to the $R_{rup}$. The rest of the variables are assumed to be as constant terms.

### 4.3. Correlation coefficient computation

After each round of simulation that is described in section 4.2, two vectors related to two distinct intensity measures are obtained for the calculation of the correlation coefficients. One of these above-mentioned intensity measures is the duration-related parameter, the significant duration. While two intensity measures of interest are independently generated by each round of simulation, it is possible to estimate the correlation coefficient using the Pearson product-moment correlation estimator:

$$\rho_{x,y} = \frac{\sum_n[(x_i - \bar{x})(y_i - \bar{y})]}{\sqrt{\sum_n[(x_i - \bar{x})^2]\sum_n[(y_i - \bar{y})^2]}} \tag{8}$$

Where $x_i$ and $y_i$ are the components of vector *X* and *Y*, which are related to the two selected intensity measures respectively; x and y are the vector means of *X* and *Y*, and $\sum_n[\ ]$ represents summation over the number of iteration or sampling in each round of simulation—or the number of the generated earthquakes in each simulation round. In this study, the number of iterations and samplings has been selected to be 5000 in each round of simulation. The correlation coefficient is a random variable which varies from one ground motion data set to another. The median correlation coefficient is determined by averaging of 10 ground motion data set. In fact, for each ground motion set type, 10 simulation procedures are conducted to generate the related ground motion data sets.

## 5. Computed correlations

In this section, the computed correlation coefficients are presented for different considered simulation cases. In all cases, the correlation of a duration-related parameter—the D5-75 metric—with one another earthquake intensity measures is explored utilizing the simulation procedure described in the previous sections. Each simulation case, as explained in the former section, represents a condition which is completely different from one another cases considered for this investigation. Hence, the characteristics of each simulation case are initially examined and explained before we go ahead to report the computed associated correlation coefficients. It is worthy to note that the computed correlation coefficients are based on the method outlined in section 4.3, so obtained correlation coefficients are the median values given the fact that the individual correlation coefficients, calculated in each round of simulation, may be associated with considerable uncertainties.

### 5.1. Random GMs (Case 1)

As stated in section 4.2, simulation case 1 representing earthquakes or events that are generated in such a way that it looks they are from a database consisting of real earthquakes happened with different characteristics. These characteristics can include factors responsible for the soil condition of the selected sites and the ones related to the source-to-site distances and faulting mechanisms as well. Therefore, these parameters are all modeled with the random variables in order to incorporate the high variability associated with the wide range of values each involved parameter may take in such a database. In this case, the VS30 parameter is modeled with a uniform random variable with the lower and upper values equal to 100 and 1000 m/s, respectively. In terms of distance metrics, $R_{rup}$ and $R_{JB}$ are uniformly sampled but with different lower and upper bounds. The source-to-site distance or $R_{rup}$ is varied between 5 to 100 km, whereas a range of values from 5 to 35 km is decided to be assigned to the $R_{JB}$ parameter. The moment magnitude is also sampled between 4 and 8 with a uniform distribution as mentioned earlier.

Figure 4 (a) and (b) show the sampled data for the calculation of correlation coefficients of D5-75 with PGA and SA at the first structural period equal to 1 sec, SA(T1=1sec). It is important to note that these figures demonstrate how correlation coefficient in one round of simulations is computed in this simulation-based framework. In other words, these correlation coefficients, which are denoted as Corr coeff$_i$ on each plot, show the correlation of D5-75 parameter with PGA and SA(T1=1sec) in one round of simulation process. As can be seen, the correlation coefficients are calculated using two vectors composed of sampled scattering data of selected intensity measures, the D5-75 parameter versus the PGA or D5-75 versus the SA(T1=1sec). For this round of simulation, we observe that the correlation coefficients of duration-related intensity measure (D5-75) with PGA and SA(T1=1sec) are 0.2038 and 0.1564, respectively. As can be readily understood, there are no significant correlation between the investigated duration-related parameter and the considered amplitude-based intensity measures, the PGA and SA at the selected period of interest. However, the results obtained from a single round of simulation may be associated with a considerable amount of uncertainties, which make them become unreliable for any statistical interpretation. Therefore, the median correlation coefficients are computed hereafter in order to reduce any possible existed uncertainties of this present simulation case.

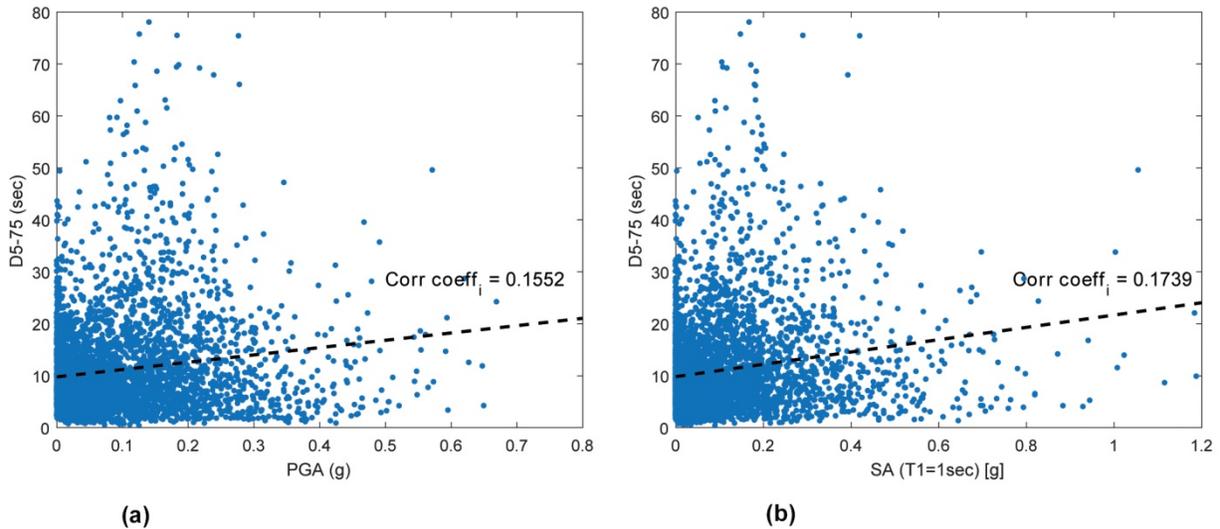

**Figure 4. The sampled data by which correlation the coefficient of two considered intensity measures is computed for a single round of simulation: a) the D5-75 versus PGA b) the D5-75 versus SA (T1=1sec)**

Figure 5 pinpoints the median correlation coefficients of the D5-75 parameter with SA for a range of applied periods of vibration of the structure. As can be found from this figure, several rounds of simulations, namely 10 times, are carried out for diminishing the potential uncertainties raised from the inherent nature of the sampling procedure. In this figure, blue curves represent the variation of correlation coefficients for each individual round of simulation, whereas the red boldface line displays the median values of correlation of D5-75 versus SA at different vibration periods. As can be seen, these two intensity measures are barely correlated during the short structural periods of vibration, for the periods between 0.02 to 0.1 second. It can be recognized for this figure that the D5-75 and SA are completely uncorrelated in structural periods around 0.11 second. However, an ascending trend for the correlation variation of D5-75 versus SA is apparently identified since the structural period of 0.11 second, which indicates that correlation of these considered intensity measures change to some extent and can increase in higher level of structural periods.

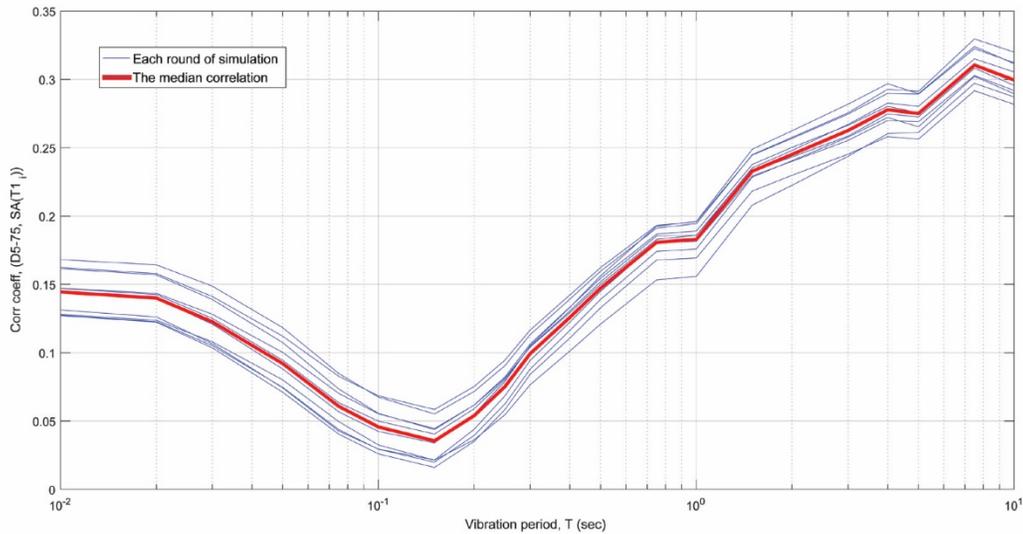

**Figure 5. The computed correlation coefficient between D5-75 parameter and SA at different structural periods of vibration, where the red line represents the median correlation coefficient. The blue lines show the variation of correlation coefficients for each round of simulation.**

Figure 6 compares the median computed correlation coefficients obtained for D5-75 parameter against different considered cumulative and amplitude-based intensity measures, the CAV, SA, PGA and PGV as well. In this figure, the vertical bars show the lower, upper and the values of correlation coefficients between these extremes, which are calculated in one of those performed rounds of simulation procedure. And the horizontal red boldface lines illustrate the median values of computed correlation coefficients that are calculated for each intensity measure against the duration-related parameter, here the D5-75. Also, the boxes on the plot are to show one standard deviation of the data. Except the SA at the higher levels of structural periods that was shown to have a correlation coefficient just about 0.3 (in figure 5), the most remarkable result to emerge from the data of figure 6 is that the duration of motion (or the D5-75 parameter) and the considered intensity measures are not generally well correlated in this simulation category.

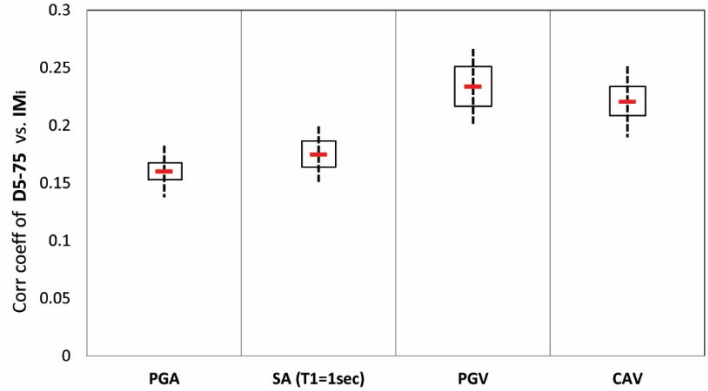

**Figure 6. The median as well as upper and lower values of the correlation coefficient between the D5-75 parameter and different considered intensity measures—the PGA, PGV, CAV and SA(T1=1sec)**

### 5.2. Same site GMs (Case 2)

The simulation case 2 can model a situation in which many produced earthquakes in a while are received by a seismometer mounted on a specific location at a considered site. It is of the essence to regard that we may have many such seismometers which are placed in different locations of where dissimilar source-to-site distances and soil conditions exist. As stated in section 4.2, just the moment magnitude is randomly sampled for this simulation case and other involved factors are considered as a constant variable. In his case, the moment magnitude, M, is modeled with a uniform distribution, where their values range from 4 to 8. As stated before, the modeling procedure associated with the selection of Ms may seem to be in contrast with the nature of the earthquake phenomenon in which larger ground shakings are rare to happen. This matter is also justifiable by the Gutenberg–Richter law. Nonetheless, a large number of simulations in high seismic levels can surely improve the reliability of our statistical analysis. Given that locations at which the seismometers are placed have a particular soil condition individually, the time-averaged shear wave velocity over a sub-surface depth 30 meters (VS30) is assumed constant and is selected to be 400 m/s. This value of VS30 is selected because it is in compliance with the soil type C as recommended by the NEHRP provisions (1997). While earthquakes are always produced and propagated from one seismic source, a range of constant rupture distances from an active fault have been chosen for this simulation example, the case 2.

Figure 7 (a) and (b) demonstrate the data trend observed in a single round of simulation prepared for the calculation of correlation coefficients of D5-75 versus PGA and SA at the first structural period equal to 1 sec, SA(T1=1sec). It is imperative to consider that these figures demonstrate how amplitude-based intensity measures and motion duration in one round of simulation case 2 are correlated to each other. For this round of simulation, we observe that the correlation coefficients of duration-related intensity measure (D5-75) with PGA and SA(T1=1sec) are 0.5556 and 0.6320, respectively. It is identified that the computed correlation coefficients in this case are relatively higher compared to the ones obtained for the simulation case 1. As can be readily found, there are good positive correlation between the duration-related parameter, D5-75 metric, and the amplitude-based intensity measures considered, the PGA and SA at the selected period of interest. However, since the results obtained from a single round of simulation may be associated with a considerable amount of uncertainties, which make our statistical analysis to become unreliable, the median correlation coefficients are computed for the rest of the results in this simulation case in order to reduce any possible involved uncertainties.

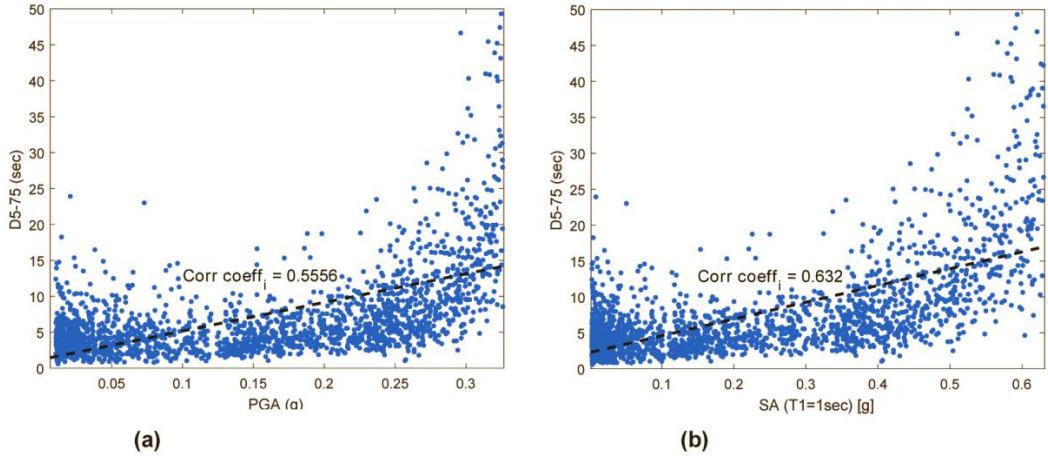

**Figure 7.** The sampled data and observed data trends by which correlation coefficient of two considered intensity measures in simulation case 2 is accomplished for a single round of simulation: a) the D5-75 versus PGA b) the D5-75 versus SA (T1=1sec)

Figure 8 has obtained through thousands of simulations and show the variation of correlation coefficients of D5-75 parameter with different cumulative and amplitude-based intensity measures (CAV, PGA, SA and PGV) as far as different values of VS30 and source-to-site distances are concerned. Figure 8 (a) and (b) show the correlation of D5-75 with PGA and SA(T1=1sec) for different VS30 parameters and increasing rupture distances. As mentioned earlier, the $R_{JB}$ is decided to be equal to 1/3 of the ones produced for increasing rupture distances. Also, the strike-slip fault is incorporated for this part of computer simulation. As can be seen from these afore-mentioned figures, small values of VS30 factor demonstrate lower values of median correlation coefficients computed for the whole range of considered rupture distances. However, the correlation coefficients are not that much affected by the variation of VS30 parameter, especially at the sites located farther from the faulting points. Thus, the median computed correlation coefficients are not heavily influenced and remained nearly unchanged for a specific opted $R_{rup}$ in terms of different VS30 parameters employed for this section of investigation. While different values of SA (or PGA) and D5-75 parameter are delivered and sampled for different taken VS30 factors as depicted in Figure 9, it is so interesting to find out that the related correlation coefficients don't show a highlighted difference. In general, the computed median correlation coefficients between the D5-75 and the amplitude-based intensity measures, the PGA or SA, get declined with regard to the large rupture distances such as the ones equal to or above 30 km (especially in figure 8 (a)). On the other hand, it can be found that motion duration and the considered amplitude-based intensity measures are well correlated at the short rupture distances. In this case, relatively high correlation coefficients, for instance, the ones just under o.65 are observed for small values of applied rupture distances.

The next two figures, namely the figure 8 (c) and (d), display the variation of correlation coefficients of the D5-75 metric with the PGV and CAV, respectively. These figures are generated based on the same simulation characteristics used in figure 8 (a) and (b)—with the same fault types as well as an identical way of producing $R_{JB}$. In both of these cases, for correlation of motion duration and PGV and CAV, the median computed correlations are slightly affected with regard to different applied VS30 parameters and reduced as larger rupture distances come about. Taken as a whole, motion duration and considered intensity measures in figures 8 (c) and (d)—the PGV and CAV—are much more correlated in the near-source earthquakes. The trends observed in these figures demonstrate that correlation of motion duration with PGV and CAV is linearly diminished regardless of which values the proposed simulation procedure employ for VS30 parameter or the variable stand for the rupture distances.

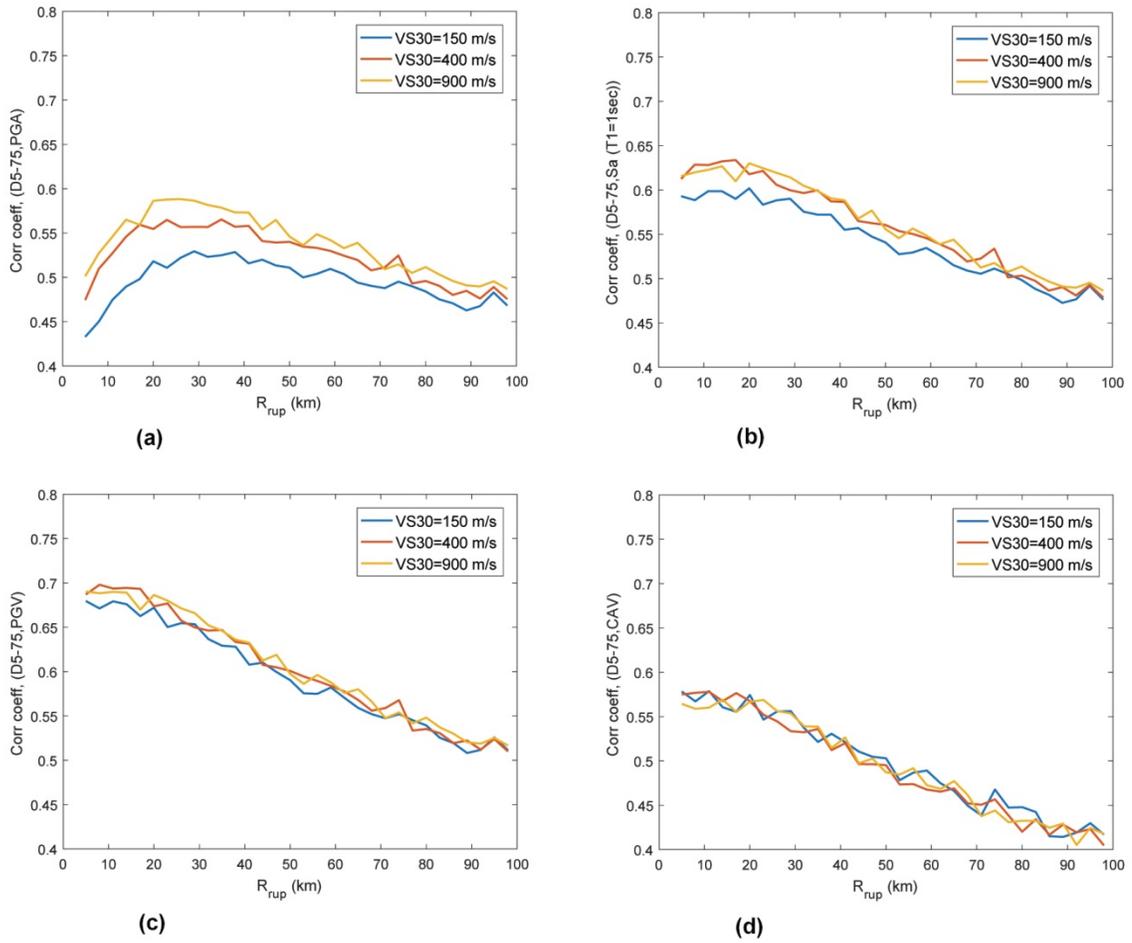

**Figure 8. Variation of the median correlation coefficient, between the D5-75 parameter and selected intensity measures, with different VS30 values and increasing rupture distances.**

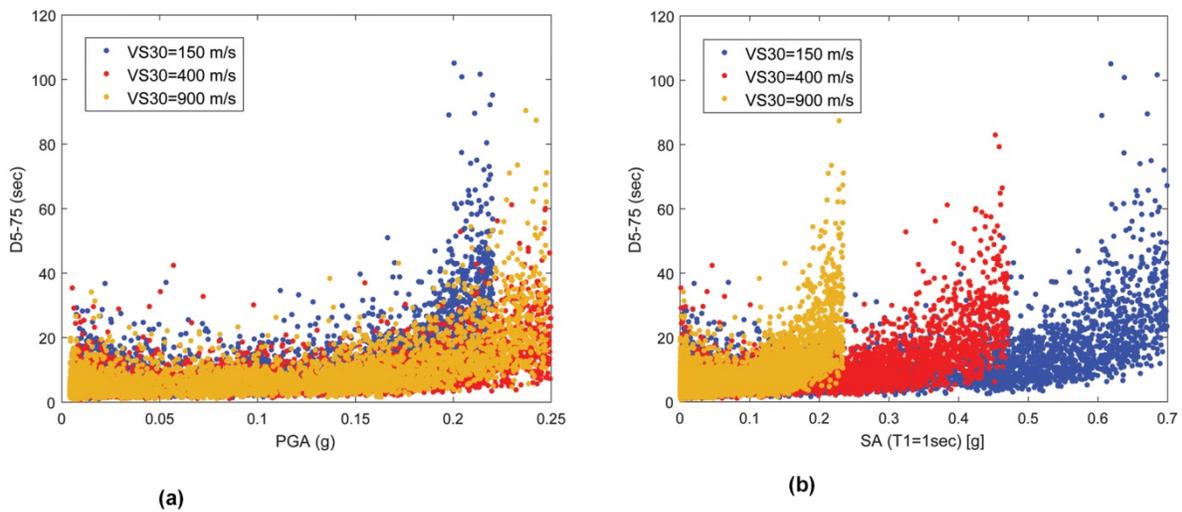

**Figure 9. Influence of VS30 factor on the data sampling of D5-75 parameter and the selected amplitude-based intensity measures: a) for D5-75 against PGA b) for D5-75 against SA (T1=1sec)**

Considering different fault types, the results obtained for the variation of correlation coefficients of D5-75 and the SA at different structural periods, $SA(T1_i)$, is depicted in Figure 10. As can be found form this figure, three faulting mechanisms have been utilized in order to check the influence of the fault types on the correlation of motion duration and the $SA(T1_i)$. As stated before, the VS30 chosen for this section of research is equal to 400 m/s based on the recommendations of the NEHRP guideline. And a 15 km rupture distance is considered in the simulation procedure, where the $R_{JB}$ is decided to be equal to 1/3 of the $R_{rup}$. The results found for normal and strike-slip faults are nearly the same for the whole range of periods of vibration, but as can be seen, the correlation coefficients of motion duration and $SA(T1_i)$ for the reverse fault are relatively below the ones computed for cases based on normal and strike-slip faulting mechanisms. Moreover, the median correlation coefficients between $SA(T1_i)$ and D5-75 parameter get steadily declined to hit a low at the vicinity of the structural period equal to 0.11 second. Consequently, except for the reverse fault that can show a clear difference, the correlation of motion duration and SA is not much affected by the faulting mechanisms.

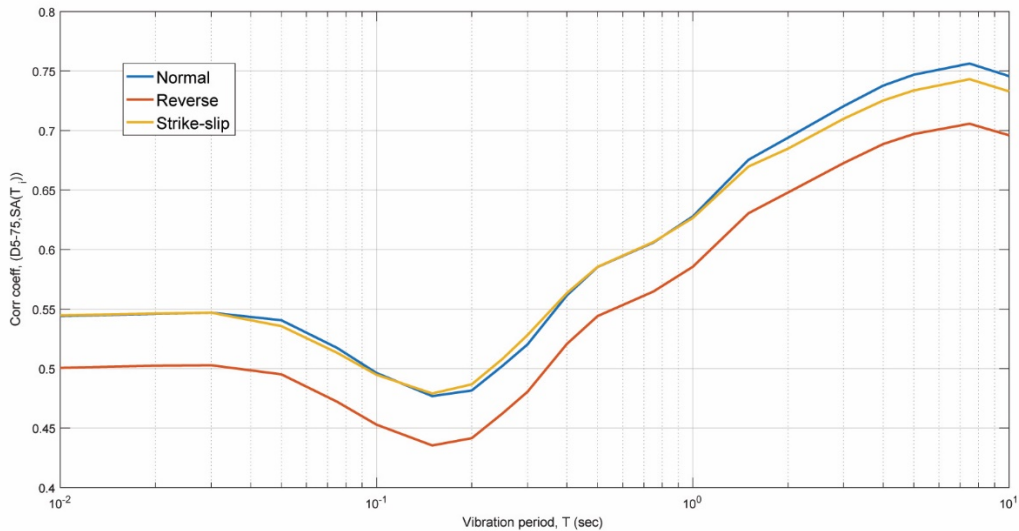

**Figure 10. The computed median values of the correlation coefficient for D5-75 parameter versus SA at various structural periods when different fault types are concerned. The fault types are the normal (blue line), reverse (red line) and the strike-slip fault (yellow line).**

### 5.3. Same source GMs (Case 3)

While in former simulation cases, many earthquakes are generated and subsequently received by the seismometers mounted on the different or particular sites, the simulation case 3 presents a condition in which just one earthquake is randomly sampled but received by several seismometers located on the paths the earthquake going through. As described in section 4.2, the parameters involved except the M—the VS30 as well as the rupture and Joyner-Boore distances—are incorporated in the simulation using random variables with the customized uniform distribution. The values associated with the VS30 parameter range from 100 to 1000 m/s. Similarly, rupture distances are randomly sampled from 5 to 100 km, but the Jonyer-Boore distance is not sampled independently and would be equal to 1/3 of those ones randomly produced for the $R_{rup}$.

The correlation of D5-75 versus the SA at different structural periods of normal fault is presented in Figure 11 when different values of Mw are applied in this simulation case. It is of worth to note that the strike-slip fault is included for figuring out the results depicted in this figure. The median correlation coefficients between motion duration (or the D5-75 parameter) and SA are computed at different hazard levels, including the hazard levels that correspond to the Mw of 4.5, 6 and 7.5. As can be readily recognized, for all values of Mw accounted, the motion duration and SA are negatively correlated at different vibration periods. However, for cases in which an Mw of 7.5 is employed, motion duration and SA are less uncorrelated though no signs of positive correlations are identified is such cases. The effect of different fault types on the correlation of motion duration and the selected amplitude-based intensity measures, the PGA and SA(T1=1sec), are also investigated in Figure 12 for the simulation case 3. As can be understood from figures 12 (a) and (b), there are no significant differences between the correlation of D5-75 parameter with PGA and SA in terms of different applied fault types—the normal, and reverse as well as strike-slip faults.

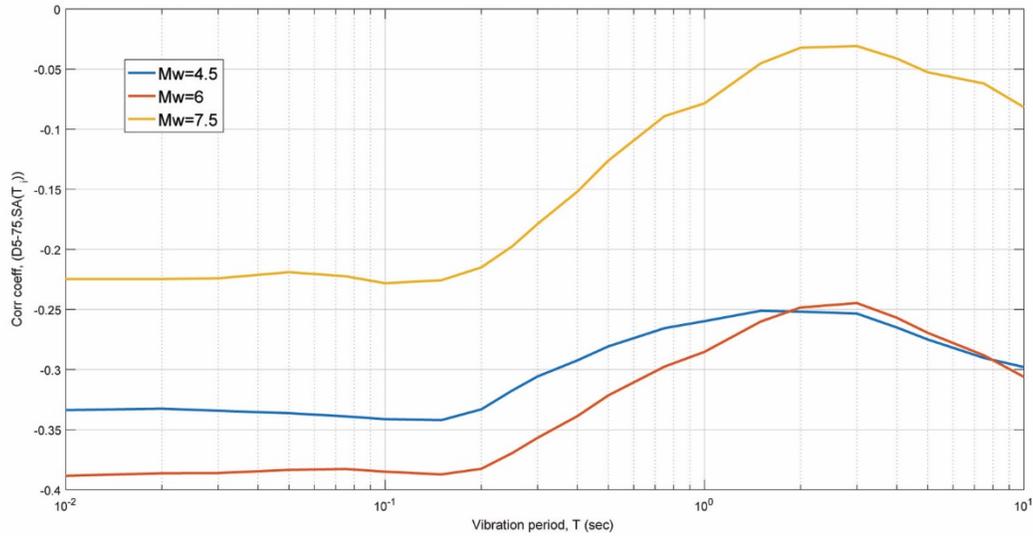

**Figure 11. The computed median values of the correlation coefficient for D5-75 parameter versus SA at various structural periods when different moment magnitude (Mw) are considered. The Mw of 4.5, 6 and 7.5 are represented by blue, red and yellow solid lines, respectively.**

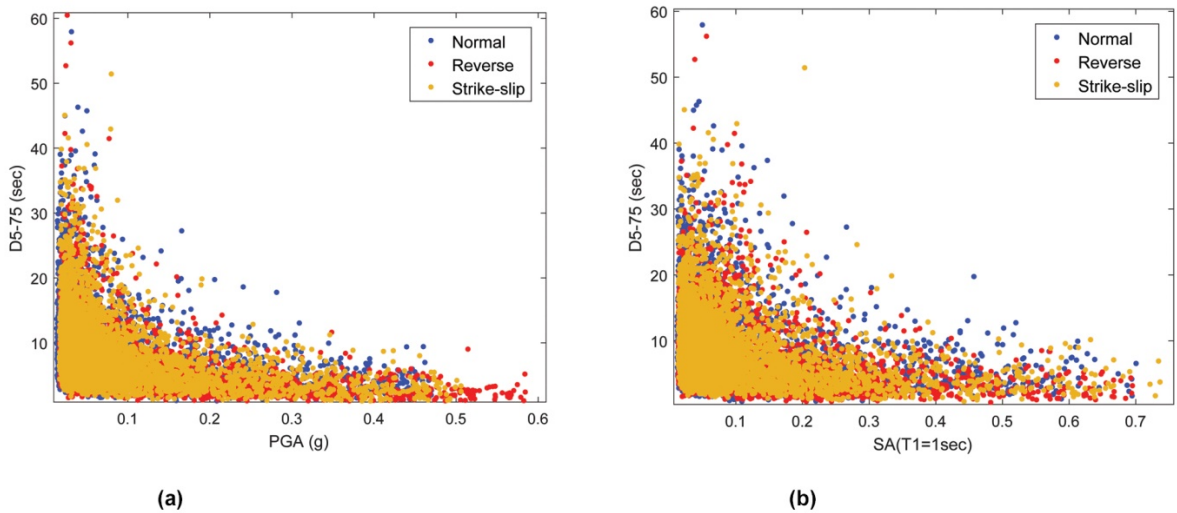

**Figure 12. Effect of fault types on the data sampling of D5-75 factor and the selected intensity measures: a) D5-75 parameter against the sampled PGA b) the D5-75 parameter against the sampled SA (T1=1sec)**

6. **Discussion**

As witnessed in the previous section, the results for correlation of motion duration with considered cumulative and amplitude-based intensity measures were often contradictory once the findings for different ground motion data sets are compared to each other. Except for the simulation case 2 that demonstrates a significant correlation of motion duration with the SA, in the next two simulation cases—the case 1 and 3—there are no signs of good correlation or even any positive correlation between duration of earthquakes and the other examined intensity measures. In this case, an utmost caution should be considered for interpreting such findings that are in substantial disagreement. The foremost cause of this discrepancy is due to the fact that our findings are based on dissimilar simulation scenarios and should not be expected to deliver the same results in fact.

In the simulation case 1 for example, the computed correlation of duration with other intensity measures is related to the events or earthquakes that are extracted from a database composed of many different earthquakes that occurred during a time window such as 20 years or more in a region of interest. Such earthquakes available from these databases are usually recorded at many different sites that may have varied characteristics. These various physical characteristics may be pertinent to the different sites that are at points with different source-to-site distances and soil conditions. Hence, the related correlation of motion duration with other intensity measures is not considered and computed for the earthquakes individually; the motion duration of an earthquake along with the amplitude-based intensity measure of one another earthquake may be considered for the correlation computation. It means that the correlations computed are not accomplished in an event-by-event manner, and they are collectively considered for the whole number of earthquakes existed in the selected database. While the computed correlations, in this case, do not reflect the relationship of motion duration and other intensity measures of a specific earthquake, it seems that there is no physical interpretation for the correlations reported for this simulation case. Therefore, the correlations—which are found commonly in the literature—that are based on such databases or simulation scenarios are somehow approximate and seem not to be highly reliable for the lack of afore-mentioned physical interpretation.

The results obtained for simulation case 3 also show an unrealistic case in which there is no positive correlation whatsoever between motion duration and other examined amplitude-based and cumulative intensity measures. The negative computed correlations can be attributed to the fact that the events incorporated for the correlation analysis are all from a single earthquake that loses its intensity (the amplitude-based or cumulative intensity) while traveling to the farthest distances. Accordingly, the correlation found in this case does not represent the relationship or variation of duration of an independent earthquake versus its other related intensity measures. However, it just shows that a reverse relation can be apparently recognized between motion duration and other intensity measures of an earthquake which is traveling through the different sites with varied soil conditions. It demonstrates that motion duration gets longer when the amplitude-based intensity measure of the motion gets weakened because of the migrating the motion signals experience or the dissipation it endures in this regard.

Contrary to what we have found in the above-mentioned simulation cases, there are relatively high positive correlations of motion duration with other opted intensity measures in the simulation case 2. This high correlation can be attributed to the well-devised simulation case that can incorporate and compute the related correlation for events that are all recorded by a particular hypothesized seismometer located at a specific site of interest. In other words, this simulation model—which appears to be quite realistic—lets us know whether durations of the motions may get increased when their amplitude-based intensity measures rise to some extent. The results associated with the above-mentioned high positive correlations of this simulation case validate that motion duration and amplitude-based intensity measures (the SA or PGA) are apparently related to each other. It means that an earthquake with a naturally generated PGA (or SA(T1)) may correspond to a distinctive significant duration. In this case, the linear scaling procedure can disturb one of the natural characteristics of the selected motion since the duration of the scaled ground motion is not altered and kept as it was before. Therefore, given the fact that linear scaling of the ground motions without any attention to their duration-related parameter can alter and damage the inherent characteristics of the real motion, it is deduced that motion duration should be also regarded as the main record selection criteria.

## 7. Summary and conclusion

In this paper, a method is proposed to examine the correlation of duration and intensity measures of ground motions. In this method, the Monte Carlo Simulation (MCS) is employed for sampling, by which continuous data that includes all possible values of the involved parameters are randomly produced and sampled. Using the developed framework, different simulation cases for reflecting varied scenario earthquakes can also be modeled to find the correlation of motion duration with the other intensity measures of interest. Correlation coefficients are investigated in three cases.

In the first case, simulated ground motions differ in terms of earthquake source parameters, site characteristics, and site-to-source distances. In the second case, which is the more realistic one, ground motions are simulated in a specific site from probable earthquake events. In the third case, which is unrealistic and for comparison only, ground motions are simulated from a specific event in different sites. In order to speculate the authenticity of the developed method, the correlation coefficients of significant duration (the D5-75) as the duration-related parameter with the amplitude-based (i.e., the PGA and SA) and cumulative (the CAV metric in this paper) intensity measures are computed. The following outcomes have been drawn for the opted and applied simulation cases:

1- Ground motions with different earthquake sources, different site parameters, and different site-to-source distance show insignificant correlation between motion duration and considered intensity measures, i.e. the PGA, PGV, SA, and CAV. On average, correlation coefficient below 0.25 is observed for the correlation of motion duration versus the afore-mentioned intensity measures.

2- Ground motions simulated in a particular site demonstrate positive correlations between motion duration and other intensity measures. Results show that correlation coefficients can reach 0.7 as far as PGV and SA are considered as the intensity measures. This simulation procedure is more realistic for engineering application and site specific design purposes.

3- Parametric studies on the motions simulated in a specific site reveal that median values of correlation coefficients decrease as source-to-site distances increase. It is also found that the local soil condition, VS30, has a little impact on the correlation coefficients. It is also found that the reverse fault mechanism delivers smaller values of correlation coefficients between the D5-75 and $SA(T1_i)$.

4- Ground motions simulated from a specific earthquake event in different sites show negative correlations once motion duration is considered versus different intensity measures. This negative correlation can be attributed to the fact that the duration of the traveling waves of an earthquake gets longer when these events are recorded at the farther rupture distances and at the same time intensity measures of ground motions decrease.

5- It is shown that the selection of earthquake ground motion data set can considerably affect the correlation coefficient results from positive correlations to the negative ones. Therefore, it can be concluded that particular attention should be paid to appropriate selection of earthquake ground motions considering the correlation between intensity and ground motion duration. The lack of adequate number of ground motion data at a particular site emphasizes the need for application of simulation procedures in generating consistent data for statistically reliable seismic analysis.

## Acknowledgment

The authors would like to thank all the efforts accomplished by the staffs in the center of High-Performance Computing (HPC) of the Sharif University of Technology for providing a robust and fast platform to run our simulation cases of this research.